\documentclass[aps,amsmath,amssymb,prl,twocolumn,preprintnumbers,superscriptaddress,nobalancelastpage]{revtex4}
\usepackage{graphicx,psfrag}
\newcommand{\LSCO}{La$_{2-x}$Sr$_x$CuO$_4$}

\newcommand{\LSCOtwelve}{La$_{1.88}$Sr$_{0.12}$CuO$_4$}

\newcommand{\LSCOsuper}{La$_{2-x}$Sr$_x$CuO$_{4 + y}$}
\newcommand{\NDLSCO}{La$_{2-x-y}$Nd$_y$Sr$_x$CuO$_4$}

\newcommand{\LBCO}{La$_{2-x}$Ba$_x$CuO$_4$}
\newcommand{\QIC}{$\mathbf{Q}_{IC}$}
\newcommand{\Br}{$\mathbf{B}(\mathbf{r})$}
\begin{document}

\title{Tuning competing orders in La$_{2-x}$Sr$_{x}$CuO$_{4}$ cuprate superconductors by the
    application of an external magnetic field}

\author{J.\ Chang}
\affiliation{Laboratory for Neutron Scattering, ETH Zurich and
Paul Scherrer Institute, CH-5232 Villigen PSI, Switzerland}

\author{Ch. Niedermayer}
\affiliation{Laboratory for Neutron Scattering, ETH Zurich and
Paul Scherrer Institute, CH-5232 Villigen PSI, Switzerland}

\author{R. Gilardi}
\affiliation{Laboratory for Neutron Scattering, ETH Zurich and
Paul Scherrer Institute, CH-5232 Villigen PSI, Switzerland}

\author{N.B. Christensen}
\affiliation{Laboratory for Neutron Scattering, ETH Zurich and
Paul Scherrer Institute, CH-5232 Villigen PSI, Switzerland}

\author{H.M. R\o nnow}
\affiliation{Laboratory for Quantum Magnetism, Ecole Polytechnique
F\'{e}d\'{e}rale de Lausanne , CH-1015 Lausanne, Switzerland}

\author{D.F.\ McMorrow}

\affiliation{London Centre for Nanotechnology and Department
of Physics and Astronomy,  University College London, London, UK}

\affiliation{ISIS Facility, Rutherford Appleton Laboratory, Chilton, Didcot OX11 0QX, UK}

\author{M. Ay}
\affiliation{Laboratory for Neutron Scattering, ETH Zurich and
Paul Scherrer Institute, CH-5232 Villigen PSI, Switzerland}

\author{J. Stahn}
\affiliation{Laboratory for Neutron Scattering, ETH Zurich and
Paul Scherrer Institute, CH-5232 Villigen PSI, Switzerland}

\author{O.\ Sobolev}
\affiliation{ BENSC Hahn-Meitner-Institut, 14109 Berlin Wannsee,
Germany}

\author{A.\ Hiess}
\affiliation{Institut Laue-Langevin, BP 156, F-38042 Grenoble, France}

\author{S.\ Pailhes}
\affiliation{Laboratory for Neutron Scattering, ETH Zurich and
Paul Scherrer Institute, CH-5232 Villigen PSI, Switzerland}

\author{C.\ Baines}
\affiliation{Laboratory for Muon Spin Spectroscopy,
Paul Scherrer Institute, CH-5232 Villigen PSI, Switzerland
            }
\author{N.\ Momono}
\affiliation{ Department of Physics, Hokkaido University - Sapporo
060-0810, Japan}

\author{M.\ Oda}
\affiliation{ Department of Physics, Hokkaido University - Sapporo
060-0810, Japan}

\author{M.\ Ido}
\affiliation{ Department of Physics, Hokkaido University - Sapporo
060-0810, Japan}

\author{J.\ Mesot}
\affiliation{Laboratory for Neutron Scattering, ETH Zurich and
Paul Scherrer Institute, CH-5232 Villigen PSI, Switzerland}

\begin{abstract}
 We report the results of a combined muon spin rotation and neutron
 scattering study on  \LSCO\ (LSCO) in the vicinity of the so-called 1/8-anomaly.
 Application of a magnetic field drives the system towards
 a magnetically ordered spin-density-wave state, which is fully
 developed at 1/8 doping. The results are discussed in terms of competition
 between antiferromagnetic and superconducting order parameters.

\end{abstract}

\date{\today}


\maketitle

Competing order parameters are a central theme in
condensed matter physics. This is especially true for the study of
high temperature superconductors (HTSC),
where superconductivity occurs upon hole doping of an
antiferromagnetic Mott insulator.
As a consequence of the competition between superconductivity (SC)
and antiferromagnetism (AF) different ground states have been
identified in the underdoped regime of La-based cuprates. Among
those are (i) d-wave SC, (ii) a disordered spin glass like state,
which coexists with SC over a broad range of doping
\cite{niedermayer98}, and (iii) a spin density wave (SDW) state
with suppressed SC around a specific hole concentration $x\approx$
1/8. This so called 1/8-anomaly was first observed in
La$_{15/8}$Ba$_{1/8}$CuO$_4$ \cite{moodenbaugh88}, where the
effect is concomitant to a structural phase transition from a
high-$T$ orthorhombic (HTO) to a low-$T$ tetragonal (LTT)
phase \cite{axeprl89}. Later, a similar anomaly was found
in the LSCO system at $x \approx 0.115$ \cite{kumagai94},
however without a structural HTO-LTT transition and without a
complete suppression of SC. A stripe model with spatial
modulations of spin and charge densities
has been suggested
to account for the incommensurate (IC) magnetic and simultaneous
charge order observed in neutron diffraction experiments on
La$_{1.48}$Nd$_{0.4}$Sr$_{0.12}$CuO$_4$ (LNSCO)
\cite{tranquada95,tranquadaprb96,christensen07}. In this model
dynamic stripe correlations of spins and holes are stabilized for
$x\sim1/8$ in the LTT phase and suppress SC.

Starting from \LSCOtwelve, there are several routes to reach the
1/8-state. One way is simply to substitute Sr with Ba
\cite{fujitaprl02} or La with Nd. An alternative is to introduce
pinning centers into the CuO$_2$-planes. $\mu$SR results on Zn
doped La$_{2-x}$Sr$_x$Cu$_{1-y}$Zn$_y$O$_4$ show an enhancement of
magnetism in the vicinity of 1/8 doping \cite{Watanabeprb02},
suggesting that small amounts of nonmagnetic impurities act as
pinning centers for dynamical stripe correlations. The observation
of similar magnetic anomalies in Zn substituted Bi-2212 and
YBa$_2$Cu$_3$O$_{6+x}$ seems to indicate that the 1/8-anomaly is
not just a specific feature of the La-based compounds, but a
general property of HTSC \cite{akoshimaprb98,koike2000}.

The subtle balance between the competing orders may also be
changed by external perturbations such as magnetic fields
\cite{katanoprb2000,lakenature02} or pressure \cite{arumugam02}.
For example, it was demonstrated that the static IC magnetic
neutron response for LSCO with doping close to 1/8 is enhanced by
the application of a magnetic field perpendicular to the
CuO$_2$-planes \cite{katanoprb2000,lakenature02}. The microscopic
mechanisms behind the 1/8-anomaly, as well as the primary cause of
the field effect remain however poorly understood.

We have therefore performed a systematic study of the competition
between the AF and SC order parameters in the vicinity of 1/8
doping. By combining $\mu$SR and neutron diffraction results
obtained on the same single crystals we show that for samples in
the 1/8 doping state the AF order is already fully developed in
zero field (ZF) and can therefore not be enhanced by the
application of an external magnetic field. A field induced
enhancement of the staggered moment is only observed in samples
with a reduced ZF magnetic response.

High quality LSCO
with $x=0.105$,
$x=0.12$,
and $x=0.145$
and LNSCO
  single crystals were
grown by the travelling solvent floating zone
method~\cite{tsfz}. The Sr content was
reassessed by measuring
the structural transition from a high-$T$
tetragonal (HTT) to a
low-$T$
orthorhombic (LTO) phase, which
vary
strongly with hole doping \cite{gilardiap02,wakimotojpsj}. The
magnetic onset temperature was determined for all samples with
both $\mu$SR ($T_f^{\mu SR}$) and neutron scattering ($T_f^{n}$).
The difference among $T_f^{\mu SR}$ and $T_f^{n}$ is due to the
different observation time scales of the two experimental
techniques. Table 1 summarizes the properties of the single
crystals used in this study.

\begin{table*}
\caption{\label{tab:table3} Compilation of $\mu$SR and neutron
scattering results on La-based compounds. $\mu_{lo}$ denotes the
local Cu moment as determined by $\mu$SR. $T^{\mu SR}_f$ and
$T^{n}_f$ are the measured time scale dependent freezing
temperatures. $\delta$ is the incommensurability. The values in
the last column indicate the structural transition temperatures.}
\begin{ruledtabular}
\begin{tabular}{ccccccccc}
 1/8 Compounds & Nominal doping& $T_c$ onset& $\mu_{lo}$&$T^{\mu SR}_f$& $T^{n}_f$&$\delta$&$T_{LTT-HTO}$\\
\hline
\NDLSCO & $x=$0.12, $y=$0.4& 7 K & 0.35(2) $\mu_B$& 50 K & 65 K&0.122(4) & 70 K  \\
   \LBCO & $x=$0.125& 5 K \cite{fujitaprb04}& 0.35 $\mu_B$\cite{saviciprb02}& 40 K\cite{saviciprl05} & 50 K\cite{fujitaprb04} &0.118 \cite{fujitaprb04}& 55 K\cite{fujitaprb04}\\
   \hline
    Compound & Nominal doping& $T_c$&$\mu_{lo}/\mu_{lo}(1/8)$& $T^{\mu SR}_f$& $T^{Neu.}_f$&$\delta$&$T_{LTO-HTT}$\\\hline
     \LSCO & $x=$0.120& 27$\pm$1.5 K & 0.51(4) & 15 K & 30 K&0.125(3)& 255 K\\
    \LSCO & $x=$0.105& 30$\pm$1.5\ K & 0.36(8) & 10 K & 25 K&0.108(2)& 290 K \\

    \LSCO & $x=$0.145& 36$\pm$1.5 K & $<$0.014 & - & -&0.13(2)& 190 K\\

\end{tabular}
\end{ruledtabular}
\end{table*}

The ZF $\mu$SR experiments were performed at the  $\pi$M3 beamline
of the Paul Scherrer Institute (PSI), which provides 100~\%
spin-polarized positive muons. They are implanted into the sample
and come to rest at interstitial lattice sites without loosing
their initial polarization. The spin of the muon then acts as a
very sensitive local magnetic probe through its precession in the
internal field \Br.
In cuprate materials the muon stopping site is close to an apical
oxygen and \Br\ arises from dipolar fields created by the
surrounding Cu$^{2+}$ moments \cite{weber90}. The neutron
scattering experiments were carried out on the cold neutron
spectrometers  RITA II  at PSI, FLEX at the Hahn-Meitner Institute
and IN14 at Institut Laue-Langevin.
The experiments were performed with a fixed incoming and final
wavevector ($k_f=k_i$=1.5 or 1.9~\AA$^{-1}$). A Be- or PG-filter
was installed before the analyzer in order to eliminate higher
order contamination.
The samples were mounted in vertical 15 Tesla cryomagnets such
that ($Q_h$,$Q_k$,0) were accessible. All measurements in an
external magnetic field $\mu_0H$ were performed after field cooling.

\begin{figure}
\begin{center}
\includegraphics[width=0.435\textwidth]{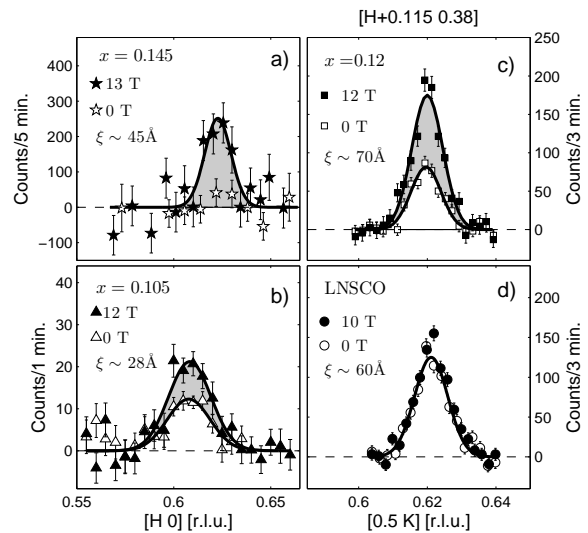}
\end{center}
\caption{ (a-d) Q-scans around \QIC\ performed on
$x=$0.145, $x=0.105$, $x=0.12$, and LNSCO respectively. Solid
lines are gaussian fits to the data. Note the different x-axis
scale for the left and right panel. The magnetic correlaton length
$\xi$ is derived from the FWHM of the magnetic peaks. Data in c
and d are resolution limited.}
\label{fig:neutron}
\end{figure}

Incommensurate AF order is observed at
$\mathbf{Q}_{IC}=(0.5,0.5\pm\delta,0),(0.5\pm\delta,0.5,0)$ in
tetragonal units of $2\pi/a = 1.65$~\AA$^{-1}$. The
incommensurability $\delta$ depends on the Sr content as in
\cite{yamada98}. Fig.~(\ref{fig:neutron}) summarizes the results
of our elastic neutron diffraction experiments. The panels in Fig.
(\ref{fig:neutron}a-d) are presented with increasing elastic
response in ZF. For $x=0.145$
 [Fig.~(\ref{fig:neutron}a)], no elastic
response is observed in ZF. However,
application of $\mu_0H=$13 T
induces an elastic response at $\delta\approx0.13$, confirming a
previous report \cite{khaykovichprb05}. For
$x=$0.105 and $x=0.12$ an elastic response  exists already in ZF
and an applied magnetic field enhances the magnetic response
for $T<T^{n}_f$ [Fig 1b,c].  LNSCO
shows the strongest ZF response and the absence of a field effect
at all $T$.

\begin{figure}
\includegraphics[width=0.41\textwidth]{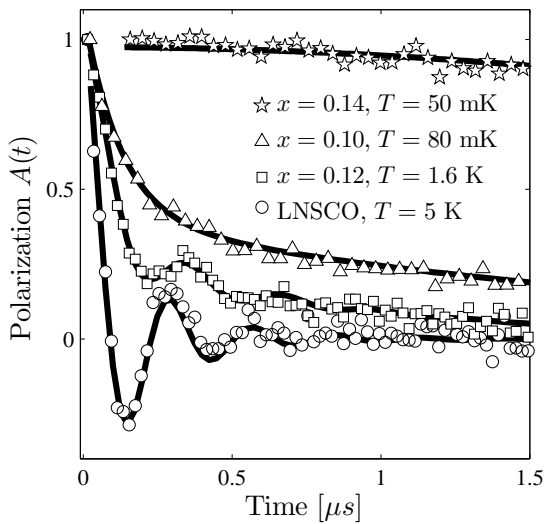}
\caption{ $\mu$SR time spectra obtained in ZF and low
temperatures. The solid lines are fits
with a Bessel function for LNSCO and $x=0.12$, a simple
exponential decay ($x=0.105$) and a Kubo-Toyabe function
($x=0.145$).} \label{fig:muon}
\end{figure}

The superconducting transition temperature is  strongly suppressed
in LNSCO
[$T_c\approx7$ K] and we therefore consider this compound to
closest mimic the physics of the so-called 1/8-state. The time
evolution of the muon spin polarization $A(t)$ exhibits a strongly
damped oscillatory behavior that is well described by a Bessel
function with a frequency $\nu\approx$3.5 MHz [see
Fig.~\ref{fig:muon}]. This observation is consistent with the
existence of an IC SDW state \cite{leprb93,saviciprb02}.

We stress that similar results are obtained for
La$_{15/8}$Ba$_{1/8}$CuO$_4$ \cite{nachumiprb98,fujitaprb04,FUAD}
and for the AF-ordered volume fraction in  superoxygenated
\LSCOsuper\ \cite{mohotallanm06}.
The latter compound was
shown to phase separate into optimally doped SC regions and an
AF-ordered phase closely related to the 1/8-state. The
characteristic features of the 1/8-state are therefore  (i) a
strongly suppressed $T_c$, (ii) the observation of a Bessel like
relaxation with $\nu\approx3.5$ MHz in the $\mu$SR time spectra
and (iii) incommensurate SDW order with $\delta\approx0.125$ and
the absence of a field effect. The 1/8-state consist, most likely,
of static stripes with an associated SDW order, which in turn
suppresses the SC order parameter. The 1/8-anomaly is limited to a
very narrow doping range and slight variations of the doping level
lead to very noticeable changes in the physical properties.

Due to its dipolar character, \Br\ is directly proportional to the
ordered Cu$^{2+}$ moment. For LNSCO the internal field at $T=5$ K
is found to be 27 mT which is about 2/3 of the value observed in
the undoped compound La$_2$CuO$_4$ \cite{uemura87,budnick87}.
Assuming a value of 0.6 $\mu_B$ for the Cu$^{2+}$ moment in
La$_2$CuO$_4$, the internal field corresponds to a local ordered
moment $\mu_{lo}\approx$0.36~$\mu_B$.

\begin{figure}
\includegraphics[width=0.4\textwidth]{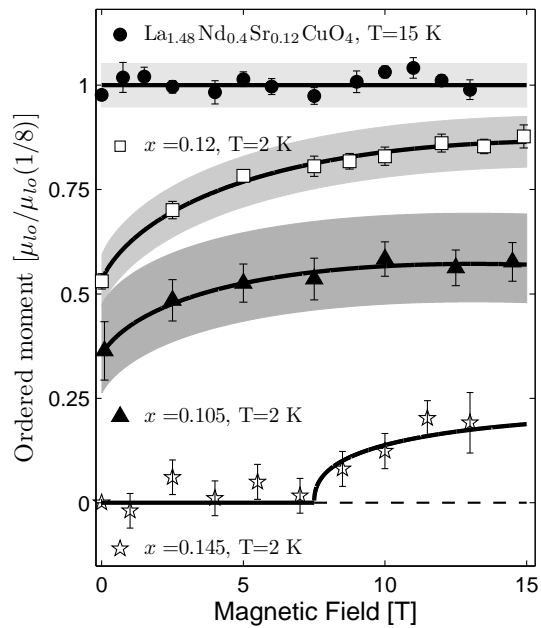}
\caption{$H$-dependence of the response at \QIC\
for LSCO $x=0.105$, $x=0.12$, $x=0.145$ and
LNSCO. The solid lines are fits to $\mu=\sqrt{I(H)}\propto
\sqrt{(H/H_c) \ln(H_c/H)}$ \cite{demler}.
Here the ZF moments estimated by $\mu$SR are used and
the gray colors indicate
the error related to the determination of the ZF order
moment. The $x=0.145$, data is presented in arbitrary units.
}
\label{fig:fig3}
\end{figure}

When moving away from 1/8 doping the ordered moment at ZF
decreases systematically with decreasing doping.
For $x=0.12$,
$A(t)$ still exhibits the characteristic Bessel type oscillation,
albeit with a reduced
frequency and an increased damping [see Fig.~\ref{fig:muon}].
Note that we observe the
full muon asymmetry, i.e. all muons experience \Br$\neq0$.
This implies that the magnetic order persists
throughout the entire volume of the sample, which is a remarkable
result for a superconductor with a $T_c$ as high as 27 K. We
emphasize that the magnetic ground state may still be
inhomogeneous but the characteristic length scale for this
inhomogeneity has to be smaller than
$\sim10-20$ \AA, which is the
typical range for dipolar fields that originate from AF ordered
moments. In fact, a nanoscale inhomogeneous state consisting of SC
droplets and patches of AF correlated regions is a
likely candidate for the ground state in a region of the phase
diagram where the antiferromagnetic and superconducting phases are
very close in energy \cite{mayrprb06,alvarezprb05}.

The neutron scattering results confirm the existence of IC
magnetic order in ZF and in addition reveal a significant
enhancement of the elastic intensity by an applied magnetic field
[see Fig.~\ref{fig:neutron}b] \cite{katanoprb2000}. Combining the
ZF $\mu$SR data with the neutron results, we are able to display
the field dependence for the different samples in a common plot
(see Fig. \ref{fig:fig3}). It can be inferred from this figure
that for $x=0.12$ the applied field tends to restore the magnetism
characteristic for 1/8 doping. This suggests that the effect of
the field is to drive the system towards the 1/8 ground state.

For $x = 0.105$, A(t) does no longer show the features of a Bessel
function, but is now well described by a single exponential
decay [see Fig.~\ref{fig:muon}]. The static nature of \Br\
was verified by
longitudinal field experiments. We deduce a static field
distribution $\Delta\approx$10 mT which is significantly reduced
from the 27 mT observed in the
1/8-compounds. Application of
$\mu_0H=12$ T doubles the amplitude of the elastic signal [see
Fig.~\ref{fig:neutron}b]. The value characteristic for static
stripe order, however, can not be fully restored in this system.

\begin{figure}
\includegraphics[width=0.5\textwidth]{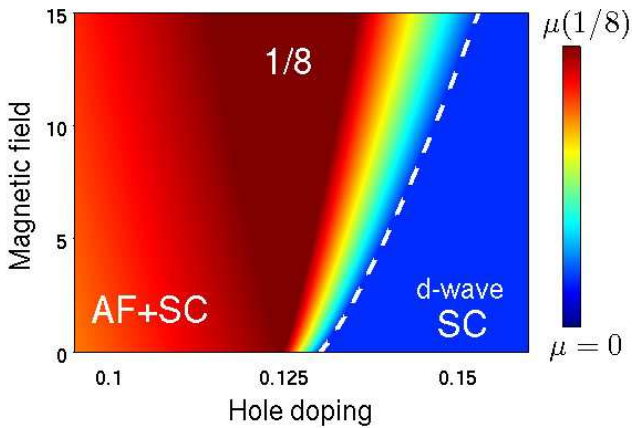}
\caption{Schematic doping-field phase diagram for \LSCO. The
ordered moment is given in false colors with red (blue) as
the maximum (minimum).
} \label{fig:fig4}
\end{figure}

For the $x = 0.145$ compound, A(t) does not exhibit any relaxation
due to electronic moments. The slow
decay of $A(t)$
is well fitted by a static Kubo-Toyabe function [see
Fig.~\ref{fig:muon}], which describes the field distribution
$P(\mathbf{B})$ arising from nuclear moments alone
\cite{hayanoprb79}. The width
of $P(\mathbf{B})$
defines an upper limit for the
electronic moments $\mu_{lo}<$0.005~$\mu_B$.
Neutron diffraction studies on this sample show field-induced
static AF-order resembling that of underdoped compounds for
$\mu_{0}H>\mu_{0}H_c$ with $\mu_{0}H_c\approx$ 7 T [see
Fig.~\ref{fig:fig4}]. A previous report found $\mu_{0}H_c\approx$
3 T \cite{khaykovichprb05}, which might indicate that the doping
level of that sample is slightly lower \cite{demler}.

We have also performed a systematic study of the vortex lattice
(VL) in $x=0.105$, $x=0.12$,
and $x=0.145$ by small angle neutron scattering.
For $x=0.145$, we observe a VL resembling that at optimum doping
\cite{Gilardi02} but only for $\mu_{0}H<6$ T. No VL could be
detected in $x=0.12$ where the largest elastic field effect is
observed. Although vortices might exist in disordered structures,
we find it difficult to correlate the elastic field effect with
vortex matter physics. Instead, we interpret our data in terms of
competing order parameters. Recently, we reported the observation
of a single d-wave gap in the ARPES spectra of the $x=0.145$
sample \cite{ming}. The most likely ZF ground state of $x=0.145$
is therefore pure d-wave SC similar to that observed at optimum
doping. Application of a magnetic field tunes the system into a
different ground state where static IC AF coexists with SC. This
ground state resembles that of more strongly underdoped LSCO
($x<0.13$) where static AF and SC orders compete even in ZF.
For $x\approx0.12$, the ordered Cu moment at $H=15$ T is close to that the 1/8 ground state [see Fig.~\ref{fig:fig3}].
Therefore we argue, that the effect of the applied field  is to drive
the system towards the 1/8 ground state.

To summarize our combined $\mu$SR and neutron diffraction
experiments we present in Fig.~\ref{fig:fig4} a schematic $H-x$
phase diagram, in which the ordered Cu moment is depicted by a
false color scheme \cite{rmk2}. The  1/8 state and the pure d-wave
SC ground state is pictured by the dark red and the blue regions,
respectively. Colors in between represent a state where AF and SC
coexist. With the application of a magnetic field one can tune the
pure SC state into the mixed state of AF and SC. At the specific
doping $x=0.12$, we found that the field drives the system towards
the 1/8 state. The different ground states are therefore very
close in energy. Our results clearly support the notion of
competing SC and static AF order parameters. The systematics of
our data shows that the existence of AF is intrinsic and not due
to defects or chemical inhomogeneities. Any suppression of
superconductivity by either a change of chemistry or by an
external pertubation goes along with a concurrent and systematic
enhancement of static magnetism.

This work was supported by the Swiss NSF (through NCCR, MaNEP, and
grant Nr 200020-105151) and the Ministry of Education and Science
of Japan. A major part of this work was
performed at the Swiss spallation source SINQ and the Swiss Muon
Source, Paul Scherrer Institut, Villigen, Switzerland.


\end{document}